\documentclass[traditabstract]{aa}  
\usepackage{natbib}
\usepackage{graphicx}
\usepackage{amsmath}
\usepackage{lipsum}
\usepackage{txfonts}
\usepackage{amssymb}
\usepackage{siunitx}

\newcommand{\Hb}{\rm H\beta }

\newcommand{\kms}{\rm km\,s^{-1}}
\newcommand{\ergs}{\rm erg\,s^{-1}}

\begin{document}

   \title{Kiloparsec-scale emission in the \\narrow-line Seyfert 1 galaxy Mrk 783}


   \author{E. Congiu
        \inst{1,2}\thanks{enrico.congiu@phd.unipd.it}
        \and M. Berton\inst{1,2}\thanks{marco.berton@unipd.it} 
        \and M. Giroletti\inst{3}
        \and R. Antonucci\inst{4}
        \and A. Caccianiga\inst{2}
        \and P. Kharb\inst{5}
        \and M. L. Lister\inst{6} \and \\       
        L. Foschini\inst{2}
        \and S. Ciroi\inst{1} 
        \and V. Cracco\inst{1}
        \and M. Frezzato\inst{1}
        \and E. J\"arvel\"a\inst{7,8}
        \and G. La Mura\inst{1}
        \and J.~L. Richards\inst{6}
        \and P. Rafanelli\inst{1}
         }

   \institute{$^{1}$ Dipartimento di Fisica e Astronomia "G. Galilei", Universit\`a di Padova, Vicolo dell'Osservatorio 3, 35122 Padova, Italy;\\
$^{2}$ INAF - Osservatorio Astronomico di Brera, via E. Bianchi 46, 23807 Merate (LC), Italy;\\ 
$^{3}$ INAF - Istituto di Radioastronomia, via Gobetti 101, Bologna I-40129, Italy;\\
 $^{4}$ Department of Physics, University of California, Santa Barbara, CA 93106-9530, USA; \\
 $^{5}$ National Centre for Radio Astrophysics - Tata Institute of Fundamental Research, Post Bag 3, Ganeshkhind, Pune 411007, India;\\
 $^{6}$ Department of Physics and Astronomy, Purdue University, 525 Northwestern Avenue, West Lafayette, IN 47907, USA;\\
 $^{7}$ Aalto University Mets{\"a}hovi Radio Observatory, Mets{\"a}hovintie 114, FIN-02540 Kylm{\"a}l{\"a}, Finland.\\
 $^8$ Aalto University Department of Electronics and Nanoengineering, P.O. Box 15500, FI-00076 AALTO, Finland.
             }

\authorrunning{E. Congiu et al.}
\titlerunning{Kiloparsec-scale radio emission in Mrk 783}

\abstract{We present the first results of a radio survey of 79 narrow-line Seyfert 1 (NLS1) carried out with the Karl G. Jansky Very Large Array (JVLA) at $5$ GHz in A configuration aimed at studying the radio properties of these sources.
We report the detection of extended emission in one object: Mrk 783.
This is intriguing, since the radio-loudness parameter R of this object is close to the threshold between radio-quiet and radio-loud active galactic nuclei (AGN).
The galaxy is one of the few NLS1 showing such an extended emission at $z<0.1$.
The radio emission is divided into a compact core component and an extended component, observed on both sides of the nucleus and extending from $14$ kpc southeast to $12$ kpc northwest.
There is no sign of a collimated jet and the shape of the extended component is similar to those of some Seyfert galaxies.
The properties of the emission are compatible with a relic produced by the intermittent activity cycle of the AGN.}

\maketitle

\section{Introduction}
Narrow-line Seyfert 1 galaxies (NLS1s) are a puzzling class of active galactic nuclei (AGN), which were first classified by \citet{Osterbrock85} according to their full width at half maximum (FWHM) of $\Hb < 2000\,\kms$. 
However, despite the narrowness of $\Hb$, their ratio [\ion{O}{III}]$\lambda5007/\Hb< 3$ and the presence of strong \ion{Fe}{II} multiplets in the optical and UV spectrum indicate that these objects are type 1 AGN. 

Radio-quiet\footnote{The radio loudness is defined by the parameter R, the ratio between the $5$ GHz flux and the optical B-band flux \citep{Kellermann89}. A source is considered to be radio-loud if $\rm R>10$ and radio-quiet if $\rm R <10$.}  NLS1s (RQNLS1s) constitute $93\%$ of the total population up to redshift $0.8$ \citep{Komossa06} and $96.5\%$ at $\rm z<0.35$ \citep{Cracco16}.
Radio-loud NLS1s (RLNLS1s) are relatively uncommon.
They can be divided into two different classes according to their radio spectrum in the cm range.
Flat-spectrum RLNLS1s (F-NLS1s) probably have a relativistic jet pointed toward Earth and can produce $\gamma$-rays \citep{Abdo09a, Abdo09b}, while steep-spectrum RLNLS1s (S-NLS1s) often show an extended radio morphology and are likely misaligned F-NLS1s. 

One of the most interesting possibilities concerning the nature of NLS1s is that they are young and evolving objects \citep{Mathur00}.
In particular, this appears to be true for RLNLS1s: F-NLS1s might be young flat-spectrum radio quasars (FSRQs) with a small black hole mass and S-NLS1s young radio galaxies \citep{Foschini15, Berton16c}.
However, a preference for low inclination might also play a role \citep[e.g.,][]{Shen14,Peterson11}.
Thus NLS1s are a somewhat heterogeneous group.

S-NLS1s have often been associated with compact steep-spectrum objects \citep[CSS;][]{Oshlack01, Komossa06, Gallo06a, Yuan08, Caccianiga14, Gu15, Schulz15, Berton16c, Caccianiga17}, which are usually believed to be young and growing radio galaxies \citep{Fanti95}. 
Only a handful of S-NLS1s were investigated in radio \citep{Whalen06, Anton08, Doi12, Richards15, Doi15, Gu15, Caccianiga17}. 
RLNLS1s indeed have a lower observed jet power than FSRQs \citep{Foschini15} because of their low black hole mass \citep{Heinz03,Foschini14}.
Therefore, while F-NLS1s are relatively easy to find because their luminosity is enhanced by relativistic beaming, S-NLS1s are not as easily detectable.

To study the radio properties of NLS1s, we carried out a survey with the Karl G. Jansky Very Large Array (JVLA) at $5$ GHz in A configuration.
Our sample consists of 60 sources drawn from the papers by \citet{Foschini15} and \citet{Berton15a}, and it contains radio-quiet (but not radio-silent) NLS1s, F-NLS1s, and S-NLS1s.
In this paper we report the detection of extended emission in one S-NLS1s, Mrk 783. 
This source is one of the few NLS1 showing such an extended emission at $\rm z<0.1$.  
In Sect.\,\ref{sec:mrk783} we describe the source according to results published in the literature, in Sect.\,\ref{sec:datared} we describe the data reduction we performed, in Sect.\,\ref{sec:results} we present our results, in Sect.\,\ref{sec:discussion} we discuss them and, finally, in Sect.\,\ref{sec:summary} we provide a brief summary. 
Throughout this work, we adopt a standard $\rm \Lambda CDM$ cosmology, with a Hubble constant $H_0 = 70\,\kms Mpc^{-1}$, and $\Omega_\Lambda = 0.73$ \citep{Komatsu11}.
Spectral indexes are specified with flux density $S_{\nu} \propto \nu^{-\alpha}$ at frequency $\nu$.

\section{Mrk 783}
\label{sec:mrk783}

Mrk\,783 (R.A. = $13$h $02$m $58.8$s Dec=$+16$d $24$m $27$s) is a NLS1 galaxy first classified by \citet{Osterbrock85} at $z = 0.0672$ \citep{Hewitt91} with a bolometric luminosity of the AGN $L_{bol} = 3.3\times10^{44}\,\ergs$ \citep{Berton15a}.
Its host galaxy was classified as a lenticular galaxy \citep{Petrosian07}, but the SDSS image clearly shows the presence of a tidal tail, or a spiral arm, extended in the east direction.

The mass of the central black hole inferred from the $\Hb$ broad component line width is about $4.3\times10^7$ M$_{\odot}$ \citep{Berton15a}.
$\Hb$ shows a prominent red wing in the broad component, indicating a receding outflow with a velocity of $\sim500\,\kms$. 
This broad component is clearly visible in all the permitted lines of the optical spectrum. 
Conversely, narrow lines, and particularly [\ion{O}{III}]$\lambda5007$, do not show any outflowing component and are well reproduced by a single Gaussian profile \citep{Berton16b}.
 
Mrk\,783 is a strong X-ray emitter that has been detected by ROSAT \citep{Schwope00}, INTEGRAL \citep{Krivonos07}, and Swift/XRT \citep{Panessa11}.  
\citet{Panessa11} reported a luminosity of $9.33\times10^{43}\,\ergs$ between $20$ and $100$ keV and a photon index of $1.7\pm0.2$ between $0.3$ and $100$ keV.
This is consistent with nonsaturated comptonization, which occurs in the accretion disk corona and not in relativistic jets.

In the last 30 years, the galaxy was observed several times in several radio bands, for example, the WSRT at $1.4$ GHz \citep{Meurs81}, VLA at $5$ GHz \citep{Ulvestad84,Ulvestad95}, and Green Bank telescope at $1.4$ GHz \citep{Bicay95}.
However, no extended emission was found.
Recently, \citet{Doi13} observed the galaxy nucleus with the Very Long Baseline Array (VLBA) looking for extended emission near the core of the AGN.
The image only shows a compact core, but the flux density recovered by the authors at $1.7$ GHz is only $4\%$ of the NRAO VLA Sky Survey (NVSS) flux density at $1.4\,$GHz \citep[$S_{\nu}=33.2\,$mJy;][]{Condon98}.
This discrepancy means that the vast majority of the flux emitted by the galaxy is distributed in structures with relatively low brightness temperature, which could not be seen by the instrument.
Another hint of the extended emission can be found in the FIRST image of the galaxy \citep{Becker95}.
The source is elongated along position angle (PA) $\ang{131}$ and shows a peak and a total flux density of $18.5$ mJy and $28.72$ mJy, respectively.
At low frequencies, the TIFR Giant Metrewave Radio Telescope Sky Survey \citep[TGSS;][]{Intema17} at $147$ MHz reports a flux density of $89.2\pm10.9$ mJy.

Mrk\,783 was classified as moderately radio-loud \citep{Berton15a} or radio-quietm \citep{Doi13}. 
The R parameter is indeed close to $10$. 
Therefore, a different estimate of the optical magnitude or optical variability in the source could have provided two different classifications. 
This is not uncommon, as has been clearly shown by \citet{Ho01} and \citet{Kharb14}.
However the radio emission does not appear to be dominant over the optical magnitude as in classical radio galaxies.

\section{Data reduction}
\label{sec:datared}

\begin{figure*}
\centering
\includegraphics[width=0.45\textwidth]{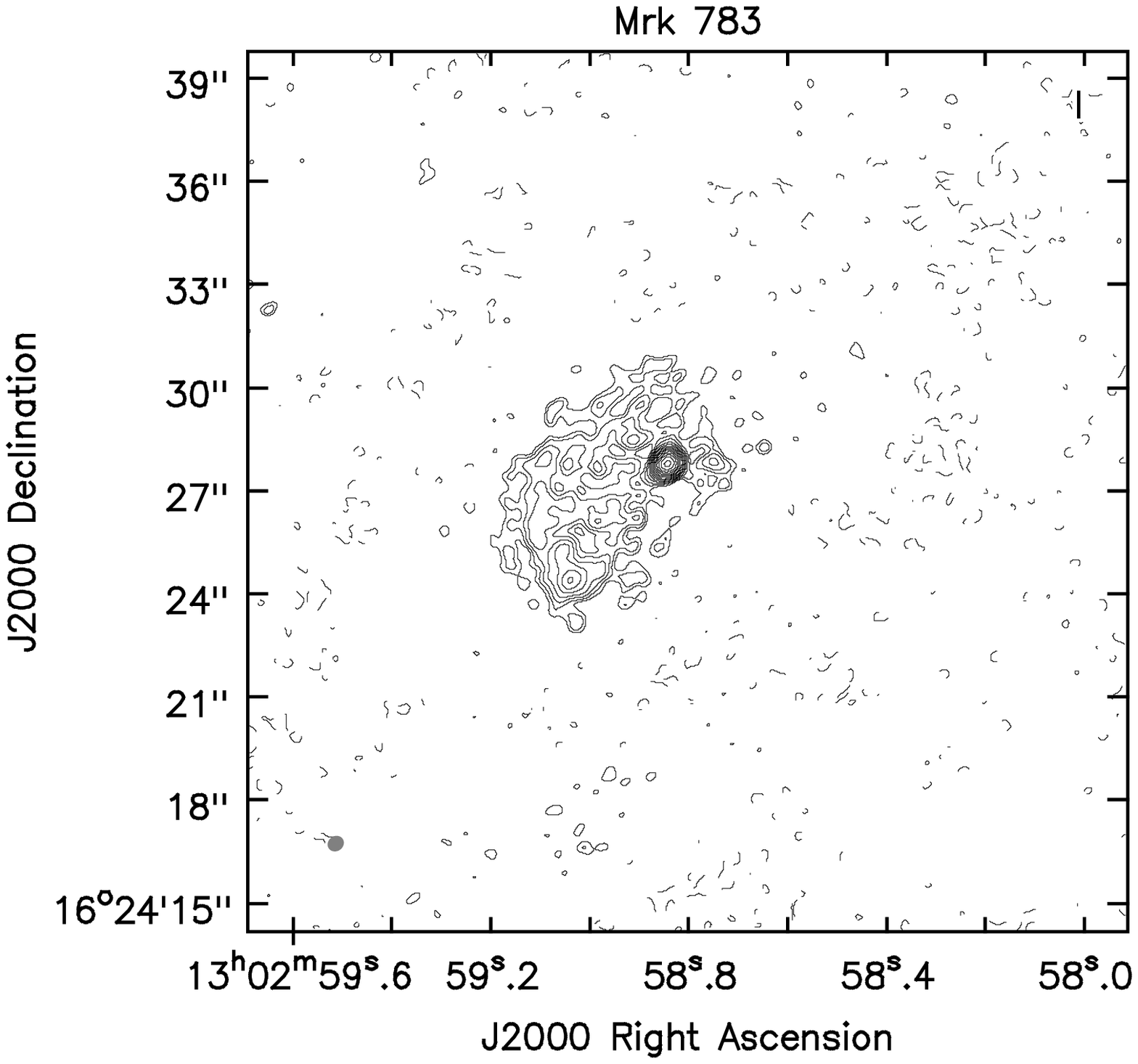} \quad
\includegraphics[width=0.45\textwidth]{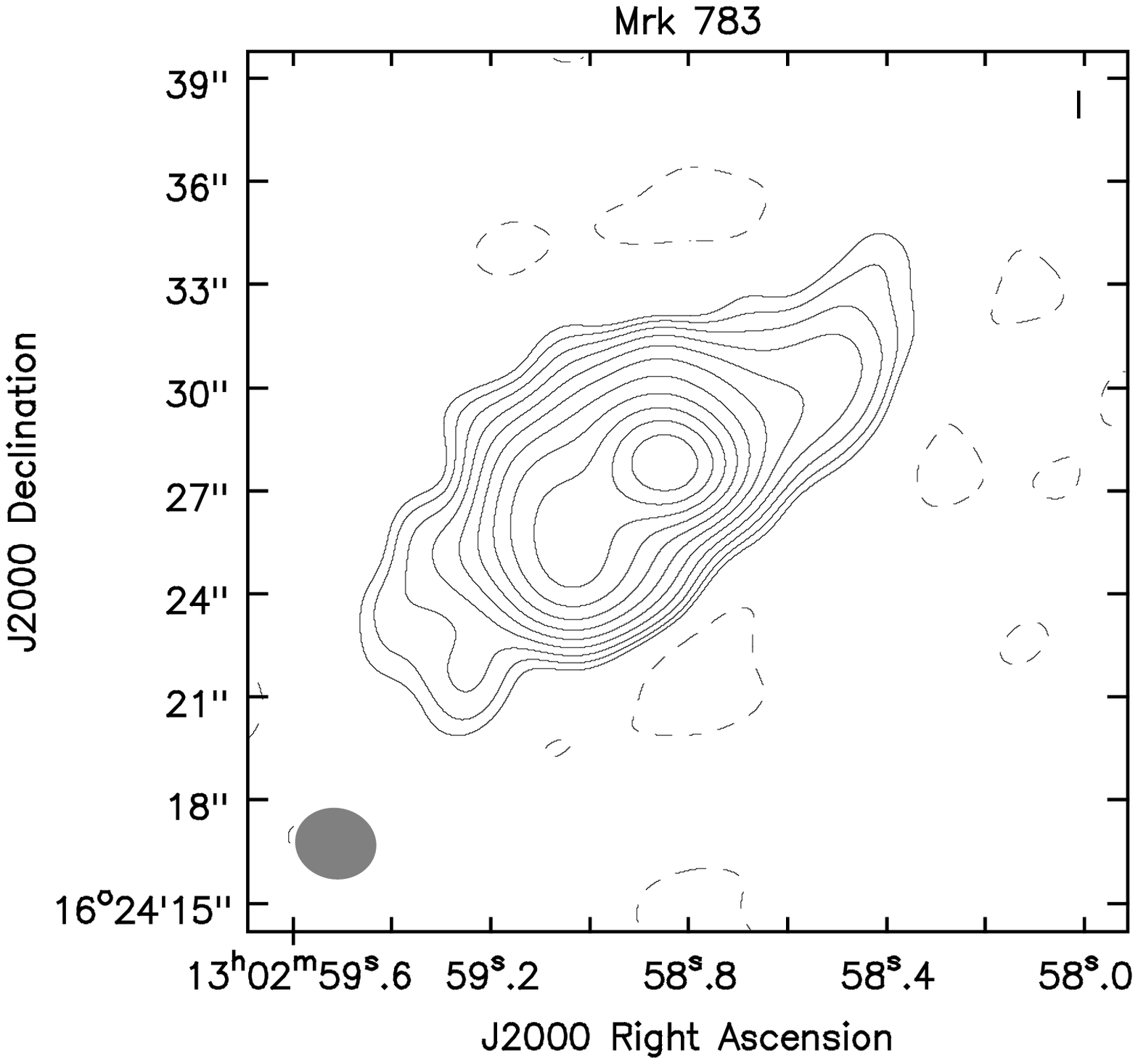}\\
\vskip-1.5cm
\qquad\qquad\includegraphics[width=0.32\textwidth]{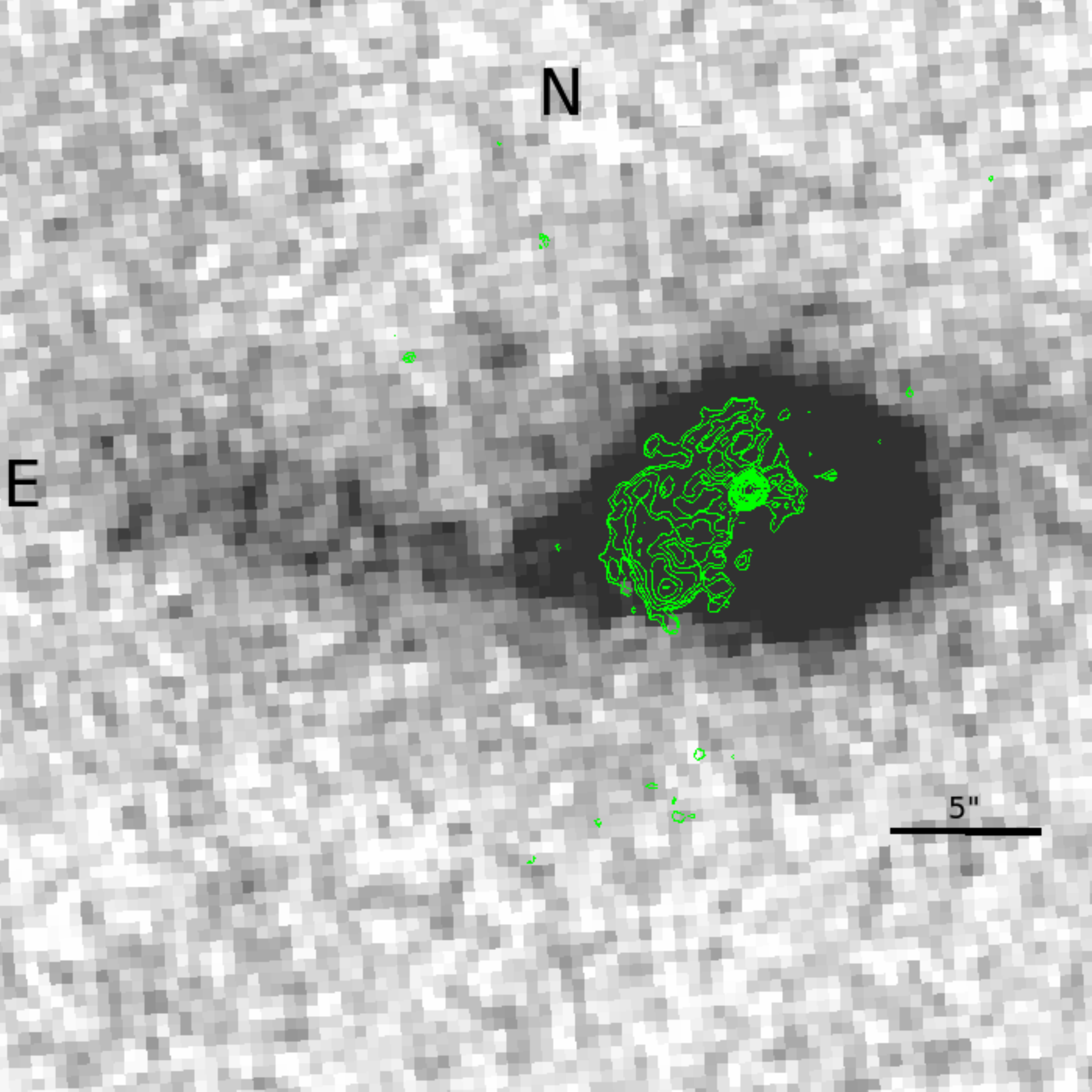} \qquad\qquad\qquad\qquad\quad
\includegraphics[width=0.32\textwidth]{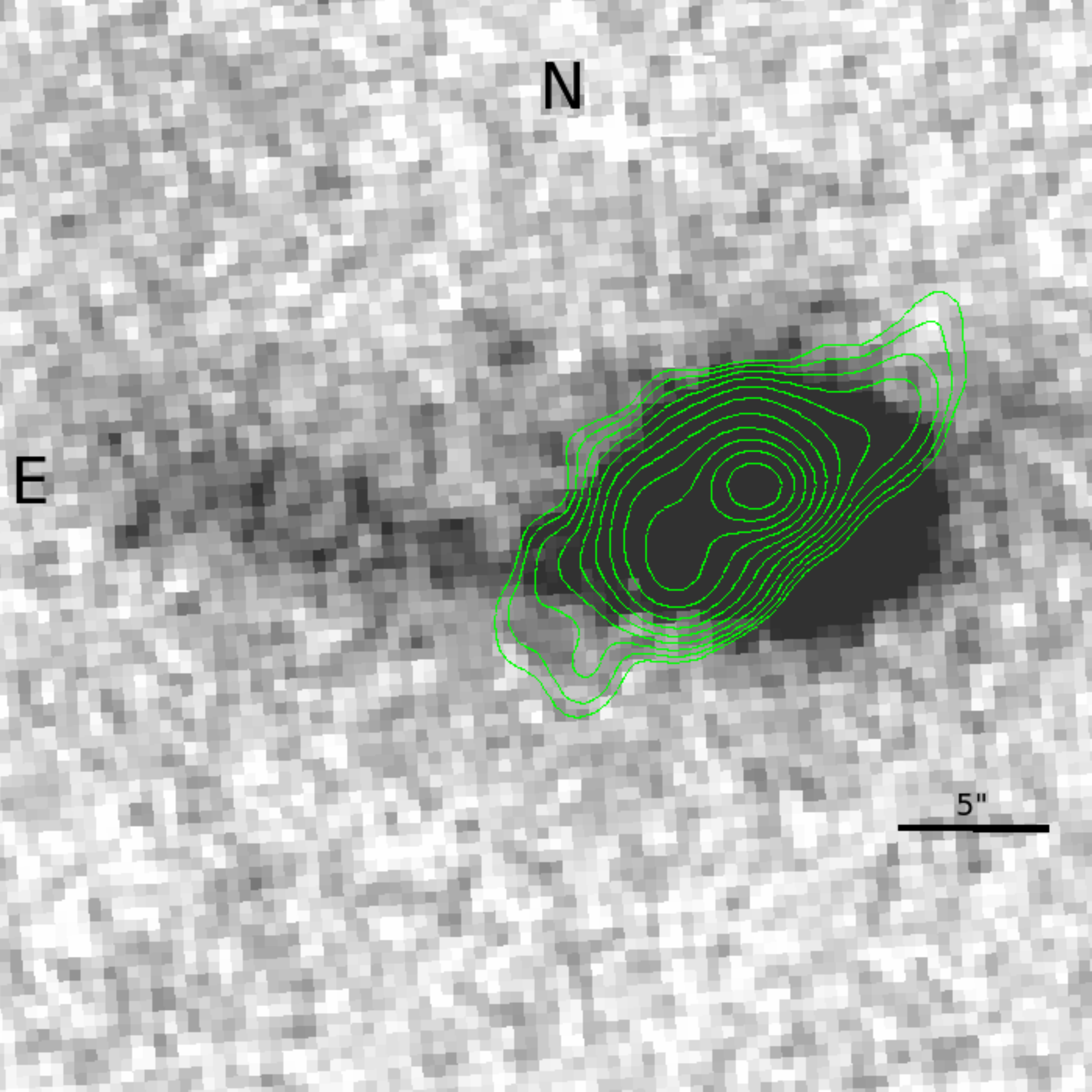}\\
\caption{\textbf{Top:} VLA images with natural weighting of Mrk\,783 at 5 GHz. The scale is $1.3$ kpc arcsec$^{-1}$. \textit{Left:} The high resolution image is shown; the beam size is 0.45"$\times$0.40". The contour levels are separated by a factor $\sqrt{2}$ starting from $3\sigma$ with $\sigma =0.011\,$mJy. The dashed contours represent a $-2\sigma$ value. \textit{Right:} Image of the source with a Gaussian taper of $50$ k$\lambda$ is shown; the beam size is 2.00"$\times$ 2.33". The contour levels are separated by a factor $\sqrt{2}$ starting from $3\sigma$ with $\sigma =0.039\,$mJy. The dashed contours represent a $-2\sigma$ value.
\textbf{Bottom:} The same contours superimposed on a SDSS image of the galaxy in the $g$ band.}
\label{fig:map}
\end{figure*}

The galaxy was observed on 2015 September 6 with the JVLA at $5$ GHz in A configuration with a bandwidth of $2$ GHz, for a total exposure time of $10$ minutes.
We reduced and analyzed the data using the Common Astronomy Software Applications (CASA) version 4.5, the standard Expanded VLA (EVLA) data reduction pipeline, and the Astronomical Image Processing System (AIPS).
The main calibrator was 3C 286. 
We split off the measurement set of the object from the main dataset, averaging over the $64$ channels of each spectral window.
After that, the object was cleaned using all the spectral windows and a natural weighting to create a first image.
To improve the quality of the final map, we proceeded with iterative cycles of phase only self-calibration of the visibilities.
The results of the CASA self-calibration were not satisfactory because this self-calibration caused a general increase of the noise level of the maps.
Therefore we tried to redo the self-calibration process using AIPS, which, indeed, significantly improved the quality of the final images.
Once the presence of extended emission was confirmed, we returned to CASA and we proceeded with a second cleaning of the data to obtain the final images.
In addition to the high resolution image we produced another image, using a taper of $50\,$k$\lambda$, to recover the highest possible fraction of the extended emission flux density.
Fig.\,\ref{fig:map} shows the maps of the object before and after the application of the taper.  
The lower panels show the radio contours superimposed on a Sloan Digital Sky Survey (SDSS) optical image.

\section{Results}
\label{sec:results}

The images in Fig.\,\ref{fig:map} clearly show a compact core and extended emission at PA$=\ang{131}$, which is in agreement with the FIRST data.
In the high resolution image the extended emission is observed only in the southeast region of the galaxy up to a projected distance of $\sim 8$ kpc.
In the tapered image, instead, we observe it on both sides of the nucleus up to a projected distance of $14$ kpc in the southeast and $12.5$ kpc in the northwest direction. 

We fitted the central component in the high resolution image with a 2D Gaussian, using the fitting algorithm in the CASA viewer.
We measured a peak flux density of $3.32\pm0.04$ mJy beam$^{-1}$, an integrated flux density $S_{\nu,c}=4.03\pm0.08$ mJy, and the size of deconvolved core axes $216\pm21$ ($\sim 280$) and $173\pm24$ ($\sim225$) mas (pc) for the major and minor axis, respectively.
The beam size is $0.45"\times0.40"$. 
With these values we calculated the core luminosity $L_{c}=2.3\times 10^{39}\,\ergs$ and we used the following equation to estimate the brightness temperature \citep{Doi13}: 
\begin{equation}
\label{eq:Tb}
T_B=1.8\times 10^9 (1+z)\frac{S_{\nu,c}}{\nu^2\phi_{maj}\phi_{min}}\sim7600 {\rm\, K},
\end{equation}
where $z$ is the redshift, $\phi_{maj}$ is the core major axis, and $\phi_{min}$ is the minor axis.
This $T_B$ is only a lower limit to the peak value because our measurement includes some extended emission.
The real core emission as detected with VLBI is smaller and \citet{Doi13} quote a peak value of $T_B>7.7\times 10^7$ K at $1.7$ GHz.
From the tapered image we measured a total flux density $S_{\nu,t} = 18.85 \pm 0.03$ mJy ($L_{tot}=1.03\times10^{40}\,\ergs$) and we obtained the flux density of the extended emission ($S_{\nu,e} = 14.82\pm0.1$ mJy, $L_{ext}=8.5\times10^{39}\,\ergs$), after subtracting the JVLA core flux density.
To evaluate the relative importance of the core emission with respect to the total emission we measured the core dominance parameter (CD) as follows:
\begin{equation}
\label{eq:cd}
{\rm CD}=S_{\nu,c}/(S_{\nu,t}-S_{\nu,c})\sim0.27.
\end{equation}
The core flux is likely an upper limit, therefore the CD parameter is also an upper limit, meaning that the galaxy is not core dominated.

To better characterize the emission we measured the spectral indexes of the two components.
We measured these components by dividing the spectral windows into two bins: the first centered at $4.7$ GHz and the second at $5.7$ GHz.
Images were made at the two frequencies, adjusting the weighting of the visibilities using the ROBUST parameter in the AIPS task IMAGR so as to obtain similar beams; they were finally convolved with identical circular beams of size $0.45$ arcsec. 
The spectral index images were made using task COMB after blanking total intensity values below $1.5$ sigma. 
We obtained the following spectral indexes for  the core component and the southern lobe, respectively:  $\alpha_{c} \sim 0.67\pm0.13$,  $\alpha_{ext} \sim 2.02\pm0.74$. 
The $\alpha_{ext}$ was estimated for the southeast lobe, avoiding the noisiest regions toward the north and east of the core.

\begin{table}
\caption{Fluxes of the galaxy in several radio and infrared bands.}
\label{tab:flux}
\centering
\begin{tabular}{lccc}
\hline
Band&Flux density (mJy) &Notes &Reference\\
\hline
$150$ MHz&$89.2$&total&1\\
$1.4$ GHz&$33.2$&total& 2\\
$1.4$ GHz&$28.7$&total& 3\\
$5$ GHz&$4.0$ & core&4\\
$5$ GHz&$14.8$ & extended&4\\
$5$ GHz&$18.9$ & total&4\\
$60{\,\rm \mu m}$& $310.0$& total &5\\
$22{\,\rm \mu m}$& $65.9$ & total& 6\\
\hline
\end{tabular}
\tablebib{(1) TGSS \citep{Intema17}; (2) NVSS \citep{Condon98}; (3) FIRST \citep{Becker95}; (4) this work; (5) IRAS \citep{Moshir90}; (6) WISE \citep{Wright10}.}
\end{table}

\section{Discussion}
\label{sec:discussion}

Our data revealed emission in a compact core as well as an extended component, which can be observed on both sides of the nucleus in the tapered image.
The luminosity at $5$ GHz is $L_{\nu}=2.1\times10^{30}\,{\rm \ergs Hz^{-1}}$, which is below the nominal $\sim 7\times10^{31}\,{\rm \ergs Hz^{-1}}$, the Fanaroff-Riley (FR) cutoff luminosity. 
The CD ($\leq 0.27$) indicates that the emission is not core dominated and the steep spectral index of the core is different from those of highly beamed radio quasars and BL Lacs ($\alpha \sim 0$), but it is more similar to that of CSS sources ($\alpha \sim 0.5$).
The extended component does not show any sign of an highly collimated jet, but it has a double-lobe shape that resembles the radio emission of some Seyfert galaxies, such as NGC\,6764 \citep{Hota06} and Mrk\,6 \citep{Kharb06}.

Extended radio surveys of normal Seyfert galaxies and LINERs \citep[e.g.,][]{Baum93,Gallimore06,Singh15} have discovered that kiloparsec-scale radio emission is not uncommon in these objects. 
The emission usually seems to have an AGN origin \citep{Gallimore06,Singh15}, but in some cases star formation (SF) can  contribute significantly \citep[e.g.,][]{Baum93}.
This could be true, indeed, in the case of NLS1 galaxies where the presence of SF (both in RQ and RL objects) has often been reported in the literature \citep[e.g.,][]{Sani11,Caccianiga15}.
The very red mid-IR colors ($W3-W4=2.6$) measured by the Wide-field Infrared Survey Explorer (WISE) of Mrk 783 and its strong emission at $60\,{\rm\mu m}$ measured by the InfraRed Astronomical Satellite (IRAS) ($310$ mJy) seem to support this hypothesis. 
In order to evaluate the impact of this possible SF on the observed radio emission, we computed the parameter $q22$, defined as\begin{equation}
\label{eq:q22}
q22 = \log (S_{22\,{\rm \mu m}}/S_{1.4\,{\rm GHz}}),
\end{equation}
where $S_{22\,{\rm \mu m}}$ and $S_{1.4\,{\rm GHz}}$ are the WISE $22\,{\rm \mu m}$ flux density and the NVSS flux density, respectively.
The resulting value ($q22\sim0.3$) is significantly lower than that usually observed in SF galaxies \citep[$q22>1$;][]{Caccianiga15}. 
This means that, even if all the observed mid-IR emissions were produced by the SF, the expected radio flux at $1.4$ GHz would be much lower than the observed flux ($\sim20$ per cent of the observed flux) . 
Considering that part of the IR emission is likely due to the AGN, we conclude that, even if present, the SF alone cannot explain the majority of the observed radio emission.

We compared our spectral indexes with results in the literature.
Using TGSS, NVSS, and our flux densities (Tab.\,\ref{tab:flux}) we recovered a spectral index $\alpha\sim 0.44$.
This is significantly flatter with respect to our measurements.
A possible explanation for this discrepancy might be the presence of a break or a cutoff at $5$ GHz.
Furthermore, biases of the surveys toward extended emissions, for example, due to short integration time, might also cause an underestimate of the spectral index.

A possible explanation for the very steep spectral indexes observed in our VLA images might be that we are observing relic emission.
This could be supported by the absence of collimated structure in our images and in the high resolution images from \citet{Doi13}.
Relic emission has already been observed in some Seyfert galaxies \citep[e.g., NGC 4235,][]{Kharb16} and they also show very steep spectra \citep[e.g.,][]{Jamrozy04,Kharb16}.

\citet{Czerny09} found that in young radio sources with high accretion rates, radiation pressure instabilities of the accretion disk can result in intermittent activity of the radio jet.
In their model, the activity phases last for $10^3-10^4$ yr and they are separated by periods of $10^4-10^6$ yr in which the radio jet is switched off.
During the cycle, the previous activity period should manifest in the form of a relic extended emission that continues its expansion until the emitting cocoon cools down and recollapses.
The maximum extension and cooling time of the cocoon depend on many factors, such as jet power and duration of the activity phase.
In the case of NLS1, both quantities are considered relatively small, resulting in a very low detection rate of such emission in this class of AGN \citep{Czerny09,Foschini15}. 
Such extended emission might appear only if the central black hole mass is in the higher part of the mass distribution \citep[which spans between $10^6$ to $10^8$ M$_{\odot}$, e.g.,][]{Cracco16}, more precisely if M$_{BH}>10^7$ M$_{\odot}$ the BH should produce the necessary jet power to make it escape from the central regions of the AGN \citep{Doi12}.
The mass of Mrk 783 black hole is $4.3\times10^7$ M$_{\odot}$ \citep{Berton15a}, therefore it belongs to the objects that in principle could produce the extended emission.
Some of the properties of the emission, such as its size and the absence of a collimated jet, might suggest that we are observing the galaxy in one of its quiescent phases not long after the switching off of the radio jet.

It is worth noting that the large scale structure of the radio emission in the tapered image has an S-like shape, which is typical of precessing radio jets \citep{Ekers78,Parma85}.
Precessing jets could arise in binary black hole systems \citep[e.g.,][]{Roos93,Romero00,Rubinur17} or result from accretion disk instabilities \citep{Pringle96,Livio97}.
In the latter case, it might be consistent with scenario of episodic activity described previously. 
The resemblance to the case of Mrk 6 is strong \citep{Kharb06}.  

Another possible explanation of why we observe such steep spectral indexes might be that there is a strong interaction between a jet and interstellar medium of the galaxy.
This could cause shocks and magnetic field amplification, leading to greater radiative losses and steep spectra.

This source might represent one of the few examples of the elusive parent population of F-NLS1s, i.e., S-NLS1s, 
Mrk 783, which lies at the edge of the RQ/RL division.
The same is true for another S-NLS1 with extended emission, Mrk 1239 \citep{Doi15}.
This might indicate that more sources of this kind could be found among RQ objects. 
Very few RQNLS1s show detectable jets or diffuse radio emission \citep{Berton16b}.
This is partly because they are by definition faint in the radio regime, but some objects that fit the RQ definition show nonthermal core emission when observed with high sensitivity \citep{Giroletti09}.

Another hypothesis regarding the parent population of F-NLS1s is based on a different assumption about NLS1s nature. 
Some authors believe that the narrowness of permitted lines is not due to the low black hole mass, but instead to a flat BLR observed pole-on \citep[e.g.,][]{Decarli08}. 
Mrk 783, however, shows extended emission on both sides of the core, hence it likely has a non-negligible inclination. 
In this case then the low black hole mass estimate should not be significantly affected by any BLR flattening, although this might happen in other objects \citep{Shen14}. 

To better understand the nature of this object, simultaneous radio observations at different frequencies are needed to study the spectral energy distribution (SED) of the emission.
In particular, it is fundamental to investigate the reason of such a difference between our spectral indexes and what we found from data in the literature.
Also images with an intermediate spatial resolution between our data and the data from \citet{Doi13} (e.g., from e-MERLIN) could be useful to investigate the presence of an intermediate scale radio jet.

\section{Summary}
\label{sec:summary}

In this paper we present the first result of a survey of NLS1s carried out with the JVLA at $5$ GHz in A configuration.
In particular we report the detection of extended emission in the S-NLS1 Mrk\,783.
We found a compact core and extended emission that, in the tapered image, is observed on both the southeast and northwest sides of the galaxy nucleus, up to a maximum projected distance of $14$ kpc.
We excluded star formation as the dominating cause of the extended emission owing to the low value of the IR-to-radio flux ratio ($q22$).
The latter, together with the morphology of the emission, indeed suggests an AGN origin.
At the same time, in the high resolution image (Fig.\,\ref{fig:map}) we could not find any sign of a large scale collimated jet and \citet{Doi13} did not find any small scale jets.
These facts, and the very steep spectral indexes that we measured, led us to hypothesize that the extended emission might be a relic and that the source might be in a quiescent period of its activity cycle, which is in agreement with the theoretical model by \citet{Czerny09} and the young age scenario for NLS1s \citep{Mathur00}. 
Further observations are needed both in radio and in other bands to confirm our results and to investigate the nature of this object  in more detail. 

\begin{acknowledgements}
The National Radio Astronomy Observatory is a facility of the National Science Foundation operated under cooperative agreement by Associated Universities, Inc. 
This research has made use of the NASA/IPAC Extragalactic Database (NED) which is operated by the Jet Propulsion Laboratory, California Institute of Technology, under contract with the National Aeronautics and  Space Administration.
This research has made use of the NASA/ IPAC Infrared Science Archive, which is operated by the Jet Propulsion Laboratory, California Institute of Technology, under contract with the National Aeronautics and Space Administration.
This publication makes use of data products from the Wide-field Infrared Survey Explorer, which is a joint project of the University of California, Los Angeles, and the Jet Propulsion Laboratory/California Institute of Technology, funded by the National Aeronautics and Space Administration.
We thank the staff of the GMRT, that made these observations possible. GMRT is run by the National Centre for Radio Astrophysics of the Tata Institute of Fundamental Research.
Funding for the Sloan Digital Sky Survey has been provided by the Alfred P. Sloan Foundation and the U.S. Department of Energy Office of Science. The SDSS web site is \texttt{http://www.sdss.org}. 
SDSS-III is managed by the Astrophysical Research Consortium for the Participating Institutions of the SDSS-III Collaboration including the University of Arizona, the Brazilian Participation Group, Brookhaven National Laboratory, Carnegie Mellon University, University of Florida, the French Participation Group, the German Participation Group, Harvard University, the Instituto de Astrofisica de Canarias, the Michigan State/Notre Dame/JINA Participation Group, Johns Hopkins University, Lawrence Berkeley National Laboratory, Max Planck Institute for Astrophysics, Max Planck Institute for Extraterrestrial Physics, New Mexico State University, University of Portsmouth, Princeton University, the Spanish Participation Group, University of Tokyo, University of Utah, Vanderbilt University, University of Virginia, University of Washington, and Yale University. 

\end{acknowledgements}

\bibliographystyle{aa}
\bibliography{./biblio}

\begin{thebibliography}{65}
\expandafter\ifx\csname natexlab\endcsname\relax\def\natexlab#1{#1}\fi

\bibitem[{{Abdo} {et~al.}(2009{\natexlab{a}}){Abdo}, {Ackermann}, {Ajello},
  {Axelsson}, {Baldini}, {Ballet}, {Barbiellini}, {Bastieri}, {Battelino}, \&
  {Baughman}}]{Abdo09a}
{Abdo}, A.~A., {Ackermann}, M., {Ajello}, M., {et~al.} 2009{\natexlab{a}},
  \apj, 699, 976

\bibitem[{{Abdo} {et~al.}(2009{\natexlab{b}}){Abdo}, {Ackermann}, {Ajello},
  {Axelsson}, {Baldini}, {Ballet}, {Barbiellini}, {Bastieri}, {Baughman},
  {Bechtol}, \& et~al.}]{Abdo09b}
{Abdo}, A.~A., {Ackermann}, M., {Ajello}, M., {et~al.} 2009{\natexlab{b}},
  \apj, 707, 727

\bibitem[{{Ant{\'o}n} {et~al.}(2008){Ant{\'o}n}, {Browne}, \&
  {March{\~a}}}]{Anton08}
{Ant{\'o}n}, S., {Browne}, I.~W.~A., \& {March{\~a}}, M.~J. 2008, \aap, 490,
  583

\bibitem[{{Baum} {et~al.}(1993){Baum}, {O'Dea}, {Dallacassa}, {de Bruyn}, \&
  {Pedlar}}]{Baum93}
{Baum}, S.~A., {O'Dea}, C.~P., {Dallacassa}, D., {de Bruyn}, A.~G., \&
  {Pedlar}, A. 1993, \apj, 419, 553

\bibitem[{{Becker} {et~al.}(1995){Becker}, {White}, \& {Helfand}}]{Becker95}
{Becker}, R.~H., {White}, R.~L., \& {Helfand}, D.~J. 1995, \apj, 450, 559

\bibitem[{{Berton} {et~al.}(2016{\natexlab{a}}){Berton}, {Caccianiga},
  {Foschini}, {Peterson}, {Mathur}, {Terreran}, {Ciroi}, {Congiu}, {Cracco},
  {Frezzato}, {La Mura}, \& {Rafanelli}}]{Berton16c}
{Berton}, M., {Caccianiga}, A., {Foschini}, L., {et~al.} 2016{\natexlab{a}},
  \aap, 591, A98

\bibitem[{{Berton} {et~al.}(2016{\natexlab{b}}){Berton}, {Foschini}, {Ciroi},
  {Cracco}, {La Mura}, {Di Mille}, \& {Rafanelli}}]{Berton16b}
{Berton}, M., {Foschini}, L., {Ciroi}, S., {et~al.} 2016{\natexlab{b}}, \aap,
  591, A88

\bibitem[{{Berton} {et~al.}(2015){Berton}, {Foschini}, {Ciroi}, {Cracco}, {La
  Mura}, {Lister}, {Mathur}, {Peterson}, {Richards}, \&
  {Rafanelli}}]{Berton15a}
{Berton}, M., {Foschini}, L., {Ciroi}, S., {et~al.} 2015, \aap, 578, A28

\bibitem[{{Bicay} {et~al.}(1995){Bicay}, {Kojoian}, {Seal}, {Dickinson}, \&
  {Malkan}}]{Bicay95}
{Bicay}, M.~D., {Kojoian}, G., {Seal}, J., {Dickinson}, D.~F., \& {Malkan},
  M.~A. 1995, \apjs, 98, 369

\bibitem[{{Caccianiga} {et~al.}(2014){Caccianiga}, {Ant{\'o}n}, {Ballo},
  {Dallacasa}, {Della Ceca}, {Fanali}, {Foschini}, {Hamilton}, {Kraus},
  {Maccacaro}, {Mack}, {March{\~a}}, {Paulino-Afonso}, {Sani}, \&
  {Severgnini}}]{Caccianiga14}
{Caccianiga}, A., {Ant{\'o}n}, S., {Ballo}, L., {et~al.} 2014, \mnras, 441, 172

\bibitem[{{Caccianiga} {et~al.}(2015){Caccianiga}, {Ant{\'o}n}, {Ballo},
  {Foschini}, {Maccacaro}, {Della Ceca}, {Severgnini}, {March{\~a}}, {Mateos},
  \& {Sani}}]{Caccianiga15}
{Caccianiga}, A., {Ant{\'o}n}, S., {Ballo}, L., {et~al.} 2015, \mnras, 451,
  1795

\bibitem[{{Caccianiga} {et~al.}(2017){Caccianiga}, {Dallacasa}, {Ant{\'o}n},
  {Ballo}, {Berton}, {Mack}, \& {Paulino-Afonso}}]{Caccianiga17}
{Caccianiga}, A., {Dallacasa}, D., {Ant{\'o}n}, S., {et~al.} 2017, \mnras, 464,
  1474

\bibitem[{{Condon} {et~al.}(1998){Condon}, {Cotton}, {Greisen}, {Yin},
  {Perley}, {Taylor}, \& {Broderick}}]{Condon98}
{Condon}, J.~J., {Cotton}, W.~D., {Greisen}, E.~W., {et~al.} 1998, \aj, 115,
  1693

\bibitem[{{Cracco} {et~al.}(2016){Cracco}, {Ciroi}, {Berton}, {Di Mille},
  {Foschini}, {La Mura}, \& {Rafanelli}}]{Cracco16}
{Cracco}, V., {Ciroi}, S., {Berton}, M., {et~al.} 2016, \mnras, 462, 1256

\bibitem[{{Czerny} {et~al.}(2009){Czerny}, {Siemiginowska}, {Janiuk},
  {Nikiel-Wroczy{\'n}ski}, \& {Stawarz}}]{Czerny09}
{Czerny}, B., {Siemiginowska}, A., {Janiuk}, A., {Nikiel-Wroczy{\'n}ski}, B.,
  \& {Stawarz}, {\L}. 2009, \apj, 698, 840

\bibitem[{{Decarli} {et~al.}(2008){Decarli}, {Dotti}, {Fontana}, \&
  {Haardt}}]{Decarli08}
{Decarli}, R., {Dotti}, M., {Fontana}, M., \& {Haardt}, F. 2008, \mnras, 386,
  L15

\bibitem[{{Doi} {et~al.}(2013){Doi}, {Asada}, {Fujisawa}, {Nagai}, {Hagiwara},
  {Wajima}, \& {Inoue}}]{Doi13}
{Doi}, A., {Asada}, K., {Fujisawa}, K., {et~al.} 2013, \apj, 765, 69

\bibitem[{{Doi} {et~al.}(2012){Doi}, {Nagira}, {Kawakatu}, {Kino}, {Nagai}, \&
  {Asada}}]{Doi12}
{Doi}, A., {Nagira}, H., {Kawakatu}, N., {et~al.} 2012, \apj, 760, 41

\bibitem[{{Doi} {et~al.}(2015){Doi}, {Wajima}, {Hagiwara}, \& {Inoue}}]{Doi15}
{Doi}, A., {Wajima}, K., {Hagiwara}, Y., \& {Inoue}, M. 2015, \apjl, 798, L30

\bibitem[{{Ekers} {et~al.}(1978){Ekers}, {Fanti}, {Lari}, \& {Parma}}]{Ekers78}
{Ekers}, R.~D., {Fanti}, R., {Lari}, C., \& {Parma}, P. 1978, \nat, 276, 588

\bibitem[{{Fanti} {et~al.}(1995){Fanti}, {Fanti}, {Dallacasa}, {Schilizzi},
  {Spencer}, \& {Stanghellini}}]{Fanti95}
{Fanti}, C., {Fanti}, R., {Dallacasa}, D., {et~al.} 1995, \aap, 302, 317

\bibitem[{{Foschini}(2014)}]{Foschini14}
{Foschini}, L. 2014, International Journal of Modern Physics Conference Series,
  28, 1460188

\bibitem[{{Foschini} {et~al.}(2015){Foschini}, {Berton}, {Caccianiga}, {Ciroi},
  {Cracco}, {Peterson}, {Angelakis}, {Braito}, {Fuhrmann}, {Gallo}, {Grupe},
  {J{\"a}rvel{\"a}}, {Kaufmann}, {Komossa}, {Kovalev}, {L{\"a}hteenm{\"a}ki},
  {Lisakov}, {Lister}, {Mathur}, {Richards}, {Romano}, {Sievers},
  {Tagliaferri}, {Tammi}, {Tibolla}, {Tornikoski}, {Vercellone}, {La Mura},
  {Maraschi}, \& {Rafanelli}}]{Foschini15}
{Foschini}, L., {Berton}, M., {Caccianiga}, A., {et~al.} 2015, \aap, 575, A13

\bibitem[{{Gallimore} {et~al.}(2006){Gallimore}, {Axon}, {O'Dea}, {Baum}, \&
  {Pedlar}}]{Gallimore06}
{Gallimore}, J.~F., {Axon}, D.~J., {O'Dea}, C.~P., {Baum}, S.~A., \& {Pedlar},
  A. 2006, \aj, 132, 546

\bibitem[{{Gallo} {et~al.}(2006){Gallo}, {Edwards}, {Ferrero}, {Kataoka},
  {Lewis}, {Ellingsen}, {Misanovic}, {Welsh}, {Whiting}, {Boller}, {Brinkmann},
  {Greenhill}, \& {Oshlack}}]{Gallo06a}
{Gallo}, L.~C., {Edwards}, P.~G., {Ferrero}, E., {et~al.} 2006, \mnras, 370,
  245

\bibitem[{{Giroletti} \& {Panessa}(2009)}]{Giroletti09}
{Giroletti}, M. \& {Panessa}, F. 2009, \apjl, 706, L260

\bibitem[{{Gu} {et~al.}(2015){Gu}, {Chen}, {Komossa}, {Yuan}, {Shen}, {Wajima},
  {Zhou}, \& {Zensus}}]{Gu15}
{Gu}, M., {Chen}, Y., {Komossa}, S., {et~al.} 2015, \apjs, 221, 3

\bibitem[{{Heinz} \& {Sunyaev}(2003)}]{Heinz03}
{Heinz}, S. \& {Sunyaev}, R.~A. 2003, \mnras, 343, L59

\bibitem[{{Hewitt} \& {Burbidge}(1991)}]{Hewitt91}
{Hewitt}, A. \& {Burbidge}, G. 1991, \apjs, 75, 297

\bibitem[{{Ho} \& {Peng}(2001)}]{Ho01}
{Ho}, L.~C. \& {Peng}, C.~Y. 2001, \apj, 555, 650

\bibitem[{{Hota} \& {Saikia}(2006)}]{Hota06}
{Hota}, A. \& {Saikia}, D.~J. 2006, \mnras, 371, 945

\bibitem[{{Intema} {et~al.}(2017){Intema}, {Jagannathan}, {Mooley}, \&
  {Frail}}]{Intema17}
{Intema}, H.~T., {Jagannathan}, P., {Mooley}, K.~P., \& {Frail}, D.~A. 2017,
  \aap, 598, A78

\bibitem[{{Jamrozy} {et~al.}(2004){Jamrozy}, {Klein}, {Mack}, {Gregorini}, \&
  {Parma}}]{Jamrozy04}
{Jamrozy}, M., {Klein}, U., {Mack}, K.-H., {Gregorini}, L., \& {Parma}, P.
  2004, \aap, 427, 79

\bibitem[{{Kellermann} {et~al.}(1989){Kellermann}, {Sramek}, {Schmidt},
  {Shaffer}, \& {Green}}]{Kellermann89}
{Kellermann}, K.~I., {Sramek}, R., {Schmidt}, M., {Shaffer}, D.~B., \& {Green},
  R. 1989, \aj, 98, 1195

\bibitem[{{Kharb} {et~al.}(2006){Kharb}, {O'Dea}, {Baum}, {Colbert}, \&
  {Xu}}]{Kharb06}
{Kharb}, P., {O'Dea}, C.~P., {Baum}, S.~A., {Colbert}, E.~J.~M., \& {Xu}, C.
  2006, \apj, 652, 177

\bibitem[{{Kharb} {et~al.}(2014){Kharb}, {O'Dea}, {Baum}, {Hardcastle},
  {Dicken}, {Croston}, {Mingo}, \& {Noel-Storr}}]{Kharb14}
{Kharb}, P., {O'Dea}, C.~P., {Baum}, S.~A., {et~al.} 2014, \mnras, 440, 2976

\bibitem[{{Kharb} {et~al.}(2016){Kharb}, {Srivastava}, {Singh}, {Gallimore},
  {Ishwara-Chandra}, \& {Ananda}}]{Kharb16}
{Kharb}, P., {Srivastava}, S., {Singh}, V., {et~al.} 2016, \mnras, 459, 1310

\bibitem[{{Komatsu} {et~al.}(2011){Komatsu}, {Smith}, {Dunkley}, {Bennett},
  {Gold}, {Hinshaw}, {Jarosik}, {Larson}, {Nolta}, {Page}, {Spergel},
  {Halpern}, {Hill}, {Kogut}, {Limon}, {Meyer}, {Odegard}, {Tucker}, {Weiland},
  {Wollack}, \& {Wright}}]{Komatsu11}
{Komatsu}, E., {Smith}, K.~M., {Dunkley}, J., {et~al.} 2011, \apjs, 192, 18

\bibitem[{{Komossa} {et~al.}(2006){Komossa}, {Voges}, {Xu}, {Mathur}, {Adorf},
  {Lemson}, {Duschl}, \& {Grupe}}]{Komossa06}
{Komossa}, S., {Voges}, W., {Xu}, D., {et~al.} 2006, \aj, 132, 531

\bibitem[{{Krivonos} {et~al.}(2007){Krivonos}, {Revnivtsev}, {Lutovinov},
  {Sazonov}, {Churazov}, \& {Sunyaev}}]{Krivonos07}
{Krivonos}, R., {Revnivtsev}, M., {Lutovinov}, A., {et~al.} 2007, \aap, 475,
  775

\bibitem[{{Livio} \& {Pringle}(1997)}]{Livio97}
{Livio}, M. \& {Pringle}, J.~E. 1997, \apj, 486, 835

\bibitem[{{Mathur}(2000)}]{Mathur00}
{Mathur}, S. 2000, \mnras, 314, L17

\bibitem[{{Meurs} \& {Wilson}(1981)}]{Meurs81}
{Meurs}, E.~J.~A. \& {Wilson}, A.~S. 1981, \aaps, 45, 99

\bibitem[{{Moshir} \& {et al.}(1990)}]{Moshir90}
{Moshir}, M. \& {et al.} 1990, in IRAS Faint Source Catalogue, version 2.0
  (1990)

\bibitem[{{Oshlack} {et~al.}(2001){Oshlack}, {Webster}, \&
  {Whiting}}]{Oshlack01}
{Oshlack}, A.~Y.~K.~N., {Webster}, R.~L., \& {Whiting}, M.~T. 2001, \apj, 558,
  578

\bibitem[{{Osterbrock} \& {Pogge}(1985)}]{Osterbrock85}
{Osterbrock}, D.~E. \& {Pogge}, R.~W. 1985, \apj, 297, 166

\bibitem[{{Panessa} {et~al.}(2011){Panessa}, {de Rosa}, {Bassani}, {Bazzano},
  {Bird}, {Landi}, {Malizia}, {Miniutti}, {Molina}, \& {Ubertini}}]{Panessa11}
{Panessa}, F., {de Rosa}, A., {Bassani}, L., {et~al.} 2011, \mnras, 417, 2426

\bibitem[{{Parma} {et~al.}(1985){Parma}, {Ekers}, \& {Fanti}}]{Parma85}
{Parma}, P., {Ekers}, R.~D., \& {Fanti}, R. 1985, \aaps, 59, 511

\bibitem[{{Peterson}(2011)}]{Peterson11}
{Peterson}, B.~M. 2011, in Narrow-Line Seyfert 1 Galaxies and their Place in
  the Universe, 32

\bibitem[{{Petrosian} {et~al.}(2007){Petrosian}, {McLean}, {Allen}, \&
  {MacKenty}}]{Petrosian07}
{Petrosian}, A., {McLean}, B., {Allen}, R.~J., \& {MacKenty}, J.~W. 2007,
  \apjs, 170, 33

\bibitem[{{Pringle}(1996)}]{Pringle96}
{Pringle}, J.~E. 1996, \mnras, 281, 357

\bibitem[{{Richards} \& {Lister}(2015)}]{Richards15}
{Richards}, J.~L. \& {Lister}, M.~L. 2015, \apjl, 800, L8

\bibitem[{{Romero} {et~al.}(2000){Romero}, {Chajet}, {Abraham}, \&
  {Fan}}]{Romero00}
{Romero}, G.~E., {Chajet}, L., {Abraham}, Z., \& {Fan}, J.~H. 2000, \aap, 360,
  57

\bibitem[{{Roos} {et~al.}(1993){Roos}, {Kaastra}, \& {Hummel}}]{Roos93}
{Roos}, N., {Kaastra}, J.~S., \& {Hummel}, C.~A. 1993, \apj, 409, 130

\bibitem[{{Rubinur} {et~al.}(2017){Rubinur}, {Das}, {Kharb}, \&
  {Honey}}]{Rubinur17}
{Rubinur}, K., {Das}, M., {Kharb}, P., \& {Honey}, M. 2017, \mnras, 465, 4772

\bibitem[{{Sani} {et~al.}(2011){Sani}, {Marconi}, {Hunt}, \&
  {Risaliti}}]{Sani11}
{Sani}, E., {Marconi}, A., {Hunt}, L.~K., \& {Risaliti}, G. 2011, \mnras, 413,
  1479

\bibitem[{{Schulz} {et~al.}(2015){Schulz}, {Kreikenbohm}, {Kadler}, {Ojha},
  {Ros}, {Stevens}, {Edwards}, {Carpenter}, {Els{\"a}sser}, {Gehrels},
  {Gro{\ss}berger}, {Hase}, {Horiuchi}, {Lovell}, {Mannheim}, {Markowitz},
  {M{\"u}ller}, {Phillips}, {Pl{\"o}tz}, {Quick}, {Tr{\"u}stedt}, {Tzioumis},
  \& {Wilms}}]{Schulz15}
{Schulz}, R., {Kreikenbohm}, A., {Kadler}, M., {et~al.} 2015, ArXiv e-prints
  [\eprint[arXiv]{1511.02631}]

\bibitem[{{Schwope} {et~al.}(2000){Schwope}, {Hasinger}, {Lehmann}, {Schwarz},
  {Brunner}, {Neizvestny}, {Ugryumov}, {Balega}, {Tr{\"u}mper}, \&
  {Voges}}]{Schwope00}
{Schwope}, A., {Hasinger}, G., {Lehmann}, I., {et~al.} 2000, Astronomische
  Nachrichten, 321, 1

\bibitem[{{Shen} \& {Ho}(2014)}]{Shen14}
{Shen}, Y. \& {Ho}, L.~C. 2014, \nat, 513, 210

\bibitem[{{Singh} {et~al.}(2015){Singh}, {Ishwara-Chandra}, {Wadadekar},
  {Beelen}, \& {Kharb}}]{Singh15}
{Singh}, V., {Ishwara-Chandra}, C.~H., {Wadadekar}, Y., {Beelen}, A., \&
  {Kharb}, P. 2015, \mnras, 446, 599

\bibitem[{{Ulvestad} {et~al.}(1995){Ulvestad}, {Antonucci}, \&
  {Goodrich}}]{Ulvestad95}
{Ulvestad}, J.~S., {Antonucci}, R.~R.~J., \& {Goodrich}, R.~W. 1995, \aj, 109,
  81

\bibitem[{{Ulvestad} \& {Wilson}(1984)}]{Ulvestad84}
{Ulvestad}, J.~S. \& {Wilson}, A.~S. 1984, \apj, 278, 544

\bibitem[{{Whalen} {et~al.}(2006){Whalen}, {Laurent-Muehleisen}, {Moran}, \&
  {Becker}}]{Whalen06}
{Whalen}, D.~J., {Laurent-Muehleisen}, S.~A., {Moran}, E.~C., \& {Becker},
  R.~H. 2006, \aj, 131, 1948

\bibitem[{{Wright} {et~al.}(2010){Wright}, {Eisenhardt}, {Mainzer}, {Ressler},
  {Cutri}, {Jarrett}, {Kirkpatrick}, {Padgett}, {McMillan}, {Skrutskie},
  {Stanford}, {Cohen}, {Walker}, {Mather}, {Leisawitz}, {Gautier}, {McLean},
  {Benford}, {Lonsdale}, {Blain}, {Mendez}, {Irace}, {Duval}, {Liu}, {Royer},
  {Heinrichsen}, {Howard}, {Shannon}, {Kendall}, {Walsh}, {Larsen}, {Cardon},
  {Schick}, {Schwalm}, {Abid}, {Fabinsky}, {Naes}, \& {Tsai}}]{Wright10}
{Wright}, E.~L., {Eisenhardt}, P.~R.~M., {Mainzer}, A.~K., {et~al.} 2010, \aj,
  140, 1868

\bibitem[{{Yuan} {et~al.}(2008){Yuan}, {Zhou}, {Komossa}, {Dong}, {Wang}, {Lu},
  \& {Bai}}]{Yuan08}
{Yuan}, W., {Zhou}, H.~Y., {Komossa}, S., {et~al.} 2008, \apj, 685, 801

\end{thebibliography}

\end{document}